\documentclass[doublecol]{epl2}
\bibstyle{eplbib}
\usepackage{bm}
\newcommand{\eqref}[1]{(\ref{#1})}

\title{Electron-electron scattering effect on spin relaxation in multi-valley nanostructures}
\shorttitle{Electron-electron scattering effect on spin relaxation in multi-valley nanostructures} %Insert here a short version of the title if it exceeds 70 characters

\author{M.M.~Glazov\inst{1} \and E.L.~Ivchenko\inst{1}}
\shortauthor{M.M.~Glazov and E.L.~Ivchenko}

\institute{
  \inst{1} Ioffe Physical-Technical Institute RAS, 194021 St.-Petersburg, Russia\\
}

\abstract{We develop a theory of effects of electron-electron collisions on
the Dyakonov-Perel' spin relaxation in multi-valley quantum wells. It is shown that
the electron-electron scattering rate which governs the spin relaxation is different
from that in a single-valley system. The theory is applied to Si/SiGe (001)-grown
quantum wells where two valleys are simultaneously populated by free carriers. The dependences of
the spin relaxation rate on temperature, electron concentration and valley-orbit splitting
are calculated and discussed. We demonstrate that in a
wide range of temperatures the electron-electron collisions can govern spin
relaxation in high-quality Si/SiGe quantum wells.}

\begin{document}

\maketitle
\section{Introduction}
Electron spin dynamics is among the most rapidly developing
branches of the modern solid state physics due to the rise of
spintronics~\cite{dyakonov_book,zutic:323}. The prospects of
spintronics which aims at the utilization of electron spin on
equal grounds with its charge in novel semiconductor devices are
related with the possibilities to create, control and manipulate
the electron spins. The understanding of microscopic mechanisms of
electron spin decoherence and relaxation is, hence, of high
importance.

The main mechanism of electron spin relaxation in bulk
semiconductors and semiconductor quantum wells (QWs) is
Dyakonov-Perel' (or precession)
mechanism~\cite{dyakonov72,dyakonov86}. It is connected with the
spin-orbit splitting of the conduction band states which acts as a
wavevector ($\bm k$) dependent effective magnetic field with the
Larmor precession frequency $\bm \Omega_{\bm k}$. Such an
effective field arises only in noncentrosymmetric systems, the
most widespread examples of them being bulk III-V semiconductors
and QWs on their base. Although bulk Si and Ge crystals possess an
inversion center, it has been demonstrated
experimentally\cite{PhysRevB.66.195315,wilamowski:035328} that the
one-side modulation-doped Si/SiGe QW structures exhibit the Rashba
effect and, in these structures, the electron spin relaxation is
governed by precession mechanism as well. Recently, a
theoretical estimation for the electron spin-orbit splitting in
Si/SiGe heterostructures have been obtained by using the empirical
tight-binding model computation~\cite{nestoklon06,nestoklon08}.

The electron spin precession in the effective magnetic field is
interrupted by the scattering events which change randomly the
electron wavevector and, hence, the direction of the spin
precession axis. Thus, the spin relaxation rate $\tau_s^{-1}$ can
be estimated as $\langle\Omega^2_{\bm k} \tau\rangle$ where
angular brackets denote the averaging over the electron ensemble
and $\tau$ is the microscopic scattering time. Hence, the spin
relaxation is slowed down by the scattering. It is evident that
any momentum scattering process such as interaction of an electron
with static impurities, interface imperfections or phonons
stabilizes the spin. It is much less obvious that the
electron-electron scattering can also suppress the Dyakonov-Perel'
spin relaxation contributing additively to
$\tau^{-1}$~\cite{glazov02,glazov04a,brand02,leyland06,wu03prb}
and making the time $\tau$ different from the momentum relaxation
time. Indeed, it does not matter whether the electron wavevector
is changed in the process of momentum scattering, due to the
cyclotron motion or as a result of collision with other
electrons~\cite{glazov02}. It is established that nothing but an
inclusion of the electron-electron scattering allows one to
describe the temperature dependence of spin relaxation rates in
high-quality GaAs QWs~\cite{leyland06}.

Here we address the electron-electron scattering effects on spin
relaxation in Si/SiGe quantum wells. Their specific feature is the
presence of several valleys [two in case of (001)-grown QWs]
populated by electrons. The Coulomb scattering cannot transfer an
electron from one valley into another although electrons from
different valleys can interact with each other. We show here that
the microscopic scattering time $\tau$ determined by
electron-electron collisions in the multi-valley band system is
different as compared with the single-valley case studied
previously. The difference is related not only to the non-equal
Fermi energies in the single-valley and multi-valley systems with
equal electron densities but also to the different screening of Coulomb
interaction in single- and multi-valley bands.

\section{Model}

\begin{figure}
\centerline{\includegraphics[width=0.4\textwidth]{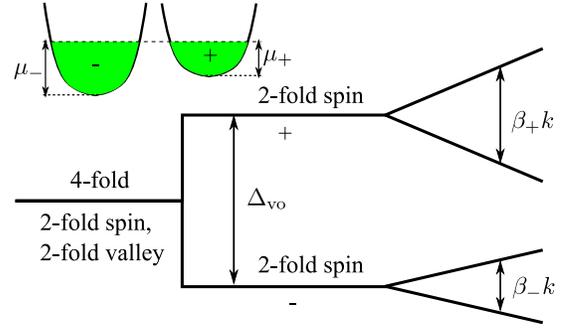}}
\caption{Schematic subband structure in an $n$-doped Si/SiGe QW. The valley-orbit
splitting, $\Delta_{\rm vo}$, and spin splitting, $\beta_+ k$ and
$\beta_- k$, are shown not to scale. Inset illustrates population of
the subbands $j = \pm$ by electrons, $\mu_+$ and $\mu_-$ are the chemical potentials referred
to the subband bottoms.}
\label{fig:splittings}
\end{figure}

To be specific we consider Si/SiGe QWs grown along the axis
$z\parallel [001]$. The conduction band states are formed from
electron states in two $\Delta$ valleys with the extrema $\pm \bm
K_0=(0,0,\pm K_0)$, where $K_0 \approx 0.8 \times 2\pi/a_0$ and
$a_0$ is the lattice constant. The electron reflection from the QW
interfaces is accompanied by the intervalley transfers $- \bm K_0
\to \bm K_0$ and vice versa which results in the valley-orbit
splitting and formation of two subbands $j = \pm$, the lower
subband $j=-$ and the higher one $j = +$. The valley-orbit
splitting $\Delta_{{\rm vo}}$, depends on the QW width and
interface properties. It may reach several meV in relatively thin
quantum wells~\cite{boykin04,nestoklon06}. The electron
eigenstates $|{\bm k}, j \rangle$ are superpositions of
single-valley states and, in the envelope-function approach, can
be written as
\begin{equation}  \label{psi}
\Psi_j ({\bm r}) = {\rm e}^{{\rm i}(k_x x + k_y y)} \hat C_s \varphi(z) [c^{(j)}_{{\bm K}_0} \psi_{{\bm K}_0}
+ c^{(j)}_{-{\bm K}_0} \psi_{-{\bm K}_0}]\:.
\end{equation}
Here $\psi_{\pm{\bm K}_0}$ are the scalar bulk Bloch functions at
the two extremum points $\pm{\bm K}_0$, $k_x, k_y$ are components
of the two-dimensional wave vector ${\bm k} \perp z$, $\varphi(z)$
is the single-valley envelope function calculated neglecting the
intervalley mixing and the spin-orbit interaction,
$c^{(j)}_{\pm{\bm K}_0}$ are coordinate independent scalar
coefficients, $\bigl|c^{(j)}_{{\bm K}_0}\bigr|^2 + \bigl|c^{(j)}_{-{\bm K}_0}\bigr|^2 = 1$, and $\hat C_s$ is a constant spinor describing the
electron spin state. In QWs with asymmetric heteropotential (or
with odd number of Si monoatomic planes) each of the subbands is
split with respect to electron spin. The typical values of the
spin-splitting have $\mu$eV range, i.e., they are much
smaller than the valley-orbit splitting. Consequently, the
electron Hamiltonian is decomposed into two partial spin-dependent
Hamiltonians
\begin{equation}
 \label{HSO}
\mathcal H^{(j)} = \frac{\hbar^2 k^2}{2 m^*} \pm \frac{\Delta_{{\rm vo}}}{2} +
\frac12 \hbar \bm \Omega_{\bm k}^{(j)} \cdot \bm \sigma \ ,
\end{equation}
describing electrons in each of the valley-orbit-split subbands.
Here $\bm \sigma$ is the vector composed of Pauli matrices and
$\bm \Omega_{\bm k}$ is the angular frequency describing the spin
splitting. The comparison of theoretical estimations and
experimental data~\cite{nestoklon08,glazov04,wilamowski07} shows
that in the state-of-the-art samples the spin splitting is
isotropic in the QW plane and has a symmetry of the Rashba type,
$\bm \Omega_{\bm k}^{(j)} = \beta_{j} (k_y, - k_x, 0)$ and
$\Omega_{\bm k}^{(j)} \equiv |\bm \Omega^{(j)}_{\bm k}| =
|\beta_{j}| k$.
The arrangement of electron states is
schematically shown in Fig.~\ref{fig:splittings}.

The kinetic theory of spin relaxation in Si/SiGe QWs is developed
within the density matrix method. It is assumed that the
valley-orbit splitting $\Delta_{\rm vo}$ can be comparable with
characteristic energy of electrons and exceeds by far the inverse
scattering time. In this case elements of the density matrix
nondiagonal in the subband indices $j \neq j'$ can be disregarded
whereas no restrictions are imposed on the density matrix in the
spin subspace. Within each subband the spin-density matrix can be
recast as
\begin{equation}
 \label{rho:gen}
\rho^{(j)}_{\bm k} = f_{\bm k}^{(j)} + \bm s_{\bm k}^{(j)} \cdot \bm \sigma \quad (j = \ \pm)\:,
\end{equation}
where $f_{\bm k}^{(j)}$ is the average occupation of the $\bm k$
state in the subband $j$, $\bm s_{\bm k}^{(j)}$ is the average
spin in this state, the symbol of the unity $2 \times 2$ matrix is
omitted.

The kinetic equation for the spin density matrix can be
represented as a set of equations for the scalar $f_{\bm k}^{(j)}$
and pseudovector $\bm s_{\bm k}^{(j)}$ as follows
\begin{eqnarray}
&&\frac{\partial f_{\bm k}^{(j)}}{\partial t} + Q_{\bm k}^{(j)}\{f,\bm s\} +
\tilde Q^{(j)}\{f,\bm s\}=0\:, \label{fk:gen} \\ \label{sk:gen}
&&\frac{\partial \bm s_{\bm k}^{(j)}}{\partial t} + \bm Q_{\bm k}^{(j)}\{\bm s,f\}
+ \tilde{\bm Q}_{\bm k}^{(j)}\{\bm s,f\} \\ &&\mbox{} \hspace{13 mm} + \bm s_{\bm k}^{(j)} \times
(\bm\Omega_{\bm k}^{(j)} + \bm \Omega_{C, \bm k}^{(j)})   = 0\:.
\nonumber
\end{eqnarray}
Here $\bm \Omega_{C,\bm k}^{(j)}$ is the effective field arising
from the Hartree-Fock interaction in the spin-polarized electron
gas~\cite{glazov04a,stich:205301}. The scalar and vector
electron-electron collision integrals, intra-valley ($Q^{(j)}_{\bm
k} \{ f,\bm s \}, {\bm Q}^{(j)}_{\bm k}\{ \bm s,f \}$) and
inter-valley ($\tilde Q^{(j)} \{f,\bm s \}$, $\tilde{\bm
Q}^{(j)}_{\bm k}\{\bm s,f\}$), are described in the next section.

\section{Intra- and inter-valley interaction}

The collision integrals in Eqs.~(\ref{fk:gen}) and (\ref{sk:gen})
describe the electron-electron scattering processes
\begin{equation} \label{collision}
(j_1 {\bm k} s_1) + (j_1^{\prime} {\bm k}^{\prime} s_1^{\prime}) \to
(j_2 {\bm p} s_2) + (j_2^{\prime} {\bm p}^{\prime} s_2^{\prime})\:,
\end{equation}
where $s_1, s_1^{\prime}$ etc. are the electron spin components
$\pm 1/2$. Because of a long-range character of the Coulomb
interaction $V_C$, the intervalley scattering accompanied by
transfer of the wavevector $\sim 2K_0$ is strongly suppressed, and
one can exclude from consideration any contributions due to the
matrix elements $\langle k'_x, k'_y, - K_0 |V_C| k_x,k_y, K_0
\rangle$ or $\langle k'_x,k'_y, K_0 |V_C| k_x,k_y, - K_0 \rangle$.

\begin{figure}
\centerline{\includegraphics[width=0.4\textwidth]{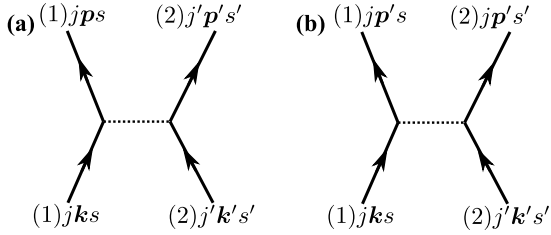}}
\caption{Illustration of the direct (a) and exchange (b) Coulomb scattering between particle 1
with the spin $s$ in the subband $j$ and particle 2 with the spin $s'$ and in the subband $j'$.
}
\label{fig:scatter}
\end{figure}

As a result, the effective matrix element describing the
process~\eqref{collision} with allowance for the
indistinguishability of the carriers reads
(c.f.~\cite{glazov04a,AlexanderPunnoose10142005}):
\begin{equation}
 \label{m:gen}
\mathcal M(j_2 {\bm p} s_2; j_2^{\prime} {\bm p}^{\prime} s_2^{\prime} | j_1 {\bm k} s_1; j_1^{\prime} {\bm k}^{\prime} s_1^{\prime}) = \delta_{\bm k+ \bm k', \bm p + \bm p'}\times
\end{equation}
$$
  \left(V_{\bm k - \bm p} \delta_{j_1j_2} \delta_{j_1'j_2'} \delta_{s_1s_2} \delta_{s_1's_2'} -V_{\bm k - \bm p'} \delta_{j_1j_2'} \delta_{j_1'j_2} \delta_{s_1s_2'} \delta_{s_1's_2} \right),
$$
where $V_{\bm k - \bm p}$ is the Fourier-transform component of
the quasi-two-dimensional Coulomb potential.
Figures~\ref{fig:scatter}(a) and \ref{fig:scatter}(b) illustrate
two contributions to the scattering process with a pair of
electrons in the final state with the wave vectors ${\bm p}$ and
${\bm p}'$. The first term in Eq.~(\ref{m:gen}) is a Coulomb
interaction where the first electron changes its wave vector from
${\bm k}$ to ${\bm p}$ while the second electron exhibits the
scattering ${\bm k}' \to {\bm p}'$. The second term results from
the scattering ${\bm k} \to {\bm p}'$ and ${\bm k}' \to {\bm p}$,
it enters Eq.~(\ref{m:gen}) with the opposite sign. In the
classical physics, the total effective cross-section is
proportional to the sum $|V_{\bm k - \bm p}|^2 + |V_{\bm k - \bm
p'}|^2$~\cite{ll1_eng}. In quantum mechanics, for two electrons which have the
same spin, $s_1=s_2$, and occupy the same subband, $j_1=j_2$,  the
cross-section has an additional interference term proportional to
$V_{\bm k - \bm p} V_{\bm k - \bm p'}$~\cite{ll3_eng}. Note, that a simple form
of the above equation stems from neglecting the spin-orbit
interaction in the processes of scattering~\cite{glazov2009}.

The collision integrals in the kinetic equations are derived by
using the standard diagram technique\cite{glazov04a} and
Eq.~\eqref{m:gen}. Here we consider the experimentally typical
situation of weak spin polarization, $|\bm s_{\bm k}^{(j)}| \ll
f^{(j)}_{\bm k}$ (although in GaAs the realization of a remarkable
optical orientation of electron spins is also
possible\cite{stich:205301}). In this case the Hartree-Fock terms
$\bm \Omega_{C, \bm k}^{(j)}$ in the kinetic
equations~\eqref{sk:gen} are unimportant and can be neglected. Let
us present the collision integrals $Q^{(j)}_{\bm k} \{ f,\bm s \}$
and $\tilde{Q}^{(j)}_{\bm k} \{f, \bm s \}$ in Eq.~(\ref{fk:gen})
in a convenient form
$$
\frac{2 \pi}{\hbar}
\sum_{ {\bm k}' {\bm p} {\bm p}'}
\delta_{{\bm k} + {\bm k}', {\bm p} + {\bm p}' }
\delta (E_k^{(j)} + E_{k'}^{(j)} - E_p^{(j)} - E_{p'}^{(j)} )
P^{(j)}_{{\bm k} {\bm k}' {\bm p} {\bm p}'}
$$
and
$$
\frac{2 \pi}{\hbar} \sum_{{\bm k}'
{\bm p} {\bm p}'}\delta_{{\bm k} + {\bm k}',\: {\bm p} + {\bm
p}'} \delta(E_k^{(j)} + E_{k'}^{(-j)} - E_p^{(j)} - E_{p'}^{(-j)})
\tilde{P}^{(j)}_{{\bm k} {\bm k}' {\bm p} {\bm p}'}\:,
$$
respectively. Here $E_k^{(j)}$ is the spin-independent part of the
electron energy equal to $\hbar^2k^2/2 m^* \pm \Delta_{\rm vo}/2$.
The above-defined scalar functions take the form
\begin{equation}\label{fs}
P^{(j)}_{{\bm k} {\bm k}' {\bm p} {\bm p}'} =
(2 V^2_{\bm k - \bm p} - V_{\bm k - {\bm p}'}V_{\bm k - \bm p})
\end{equation}
$$
\times [f_{\bm k}^{(j)}f_{{\bm k}'}^{(j)}(1-f_{\bm p}^{(j)}-f_{{\bm p}' }^{(j)}) -
f_{\bm p}^{(j)}f_{{\bm p}' }^{(j)}(1-f_{\bm k}^{(j)}-f_{{\bm k}' }^{(j)})]\:,
$$
for the intra-subband scattering, and
\begin{equation}\label{fs:inter}
\tilde{P}^{(j)}_{{\bm k} {\bm k}' {\bm p} {\bm p}'} =  2V_{\bm k - \bm p}^2
\end{equation}
$$
\times [f_{\bm k}^{(j)}f_{{\bm k}'}^{(-j)}(1-f_{\bm p}^{(j)}-f_{{\bm p}' }^{(-j)}) -
f_{\bm p}^{(j)}f_{{\bm p}' }^{(-j)}(1-f_{\bm k}^{(j)}-f_{{\bm k}' }^{(-j)})]\:,
$$
for the subband-subband scattering, similarly to the case of
electron-hole scattering and electron-electron scattering in a
quantum well with several occupied size-quantized subbands
\cite{dur96}. It is worth mentioning that for the scattering
between different particles (e.g. electrons and ions in plasma)
the scattering rates are by the factor of $2$ smaller as compared
with those given by Eq.~\eqref{fs:inter} because of the absence of
the contribution given by Fig.~\ref{fig:scatter}(b).

For the pseudovector collision integrals $\tilde{{\bm Q}}^{(j)}
\{\bm s, f \}$ and $\tilde{\bm Q}^{(j)}_{\bm k}\{\bm s, f\}$, we
similarly introduce the pseudovectors ${\bm P}^{(j)}_{{\bm k} {\bm
k}' {\bm p} {\bm p}'}$ and $\tilde{ {\bm P}}^{(j)}_{{\bm k} {\bm
k}' {\bm p} {\bm p}'}$ which are given, respectively, by
\begin{eqnarray}
&(2 V_{\bm k - \bm p}^2 - V_{\bm k - \bm p}V_{\bm
k - {\bm p}'}) [ {\bm s}_{\bm k}^{(j)} F_j({\bm k}';{\bm p},{\bm p}')
- {\bm s}_{\bm p}^{(j)} F_j({\bm p}';{\bm k},{\bm k}') ] \nonumber\\
&-  V_{\bm k - \bm p} V_{\bm k - \bm p'} [ {\bm s}_{\bm k'}^{(j)}
F_j({\bm k};{\bm p},{\bm p}') - {\bm s}_{\bm p}^{(j)} F_j({\bm p}';{\bm
k},{\bm k}')]  \:, \nonumber
\end{eqnarray}
\begin{equation}
\label{sf:inter}
2V_{\bm k - \bm p}^2 \left[ {\bm s}_{\bm k}^{(j)} \tilde F_j({\bm
k}';{\bm p},{\bm p}') - {\bm s}_{\bm p}^{(j)} \tilde F_j({\bm
p}';{\bm k},{\bm k}') \right]\:,
\end{equation}
where $F_j({\bm k}_1;{\bm k}_2,{\bm k}_3) = f_{{\bm k}_1}^{(j)} (1
- f_{{\bm k}_2}^{(j)} - f_{{\bm k}_3}^{(j)}) + f_{{\bm k}_2}^{(j)}
f_{{\bm k}_3}^{(j)}$ and $ \tilde F_j({\bm k}_1;{\bm k}_2,{\bm
k}_3) = f_{{\bm k}_1}^{(-j)} (1 - f_{{\bm k}_2}^{(j)} - f_{{\bm
k}_3}^{(-j)}) + f_{{\bm k}_2}^{(j)} f_{{\bm k}_3}^{(-j)}$. Similar
collision integrals for subband-subband scattering were derived in
Ref.~\cite{wu08} for the electron-electron collisions in GaAs
quantum well with $\Gamma$ and $L$ occupied valleys.

Before turning to the spin relaxation times we discuss the
screening of Coulomb potential in a multivalley system. Assuming
that the QW width is small enough to permit the electrons to be
treated as strictly two-dimensional, the Fourier transform of
Coulomb potential may be written approximately as, e.g., Refs.
\cite{dur96,ando},
\begin{equation}\label{vq}
 V_{\bm q} = \frac{2\pi e^2}{S \ \ae (q+q_s)}\:,
\end{equation}
where $e$ is the elementary charge, $S$ is the normalization area,
$\ae$ is the static dielectric constant, and $q_s$ is the inverse
screening length given by
\begin{equation}
 \label{qs}
q_s = \frac{2 m^* e^2}{ \mbox{\ae} \hbar^2} \sum_j \left(1 + {\rm e}^{- \mu_j / k_B
T}\right)^{-1}.
\end{equation}
Here the summation is carried out over occupied subbands, $k_B$ is
Boltzmann's constant, $T$ is the absolute temperature, $\mu_j$ is
the chemical potential of electrons referred to the bottom of the
$j$-th subband, see inset in Fig.~\ref{fig:splittings}. In the
limit of non-degenerate electrons, $\exp{(- \mu_j / k_B T )} \gg
1$, and the screening is negligible. If electrons are strongly
degenerate, $\exp{(- \mu_j/k_B T)} \ll 1$, each occupied subband
yields the same contribution $2m^*e^2/(\ae \hbar^2)$ and the total
inverse screening length increases proportionally to the number of
occupied subbands.

\section{Spin relaxation times}
Kinetic equations~\eqref{fk:gen}, \eqref{sk:gen} are solved
following the standard procedure~\cite{glazov04a}. We consider
the equilibrium electron distribution with $f_{\bm k}^{(j)}
= \{  \exp{[(E^{(j)}_k - \mu_j)/k_B T}] + 1\}^{-1}$ and seek the
spin distribution function $\bm s_{\bm k}^{(j)}$ in the form
\begin{equation}
\bm s_{\bm k}^{(j)} = \bar{\bm s}_{k}^{(j)} + \delta \bm s_{\bm k}^{(j)}\:.
\end{equation}
Here $\bar{\bm s}_{k}^{(j)}$ is a quasi-equilibrium
axially-symmetric spin distribution function related to the
initially created total electron spin in the $j$-th subband by
$\bm S^{(j)} = \sum_{\bm k} \bar{\bm s}_{k}^{(j)}$, and $\delta
\bm s_{\bm k}^{(j)}$ is a non-equilibrium correction resulting
from the electron spin precession around the vector $\bm
\Omega_{\bm k}^{(j)}$. Below we assume $\Omega_{\bm k}^{(j)} \tau
\ll 1$ (the collision dominated regime) where $\tau$ is the
typical scattering time. This condition is surely satisfied in
Si/SiGe QWs~\cite{PhysRevB.66.195315,wilamowski:035328,glazov04}.
Since the collision integrals $\bm Q_{\bm k}^{(j)}\{\delta \bm
s,f\}$ and $\tilde{\bm Q}_{\bm k}^{(j)}\{\delta \bm s,f\}$
conserve the angular dependence of $\delta \bm s_{\bm k}^{(j)}$
one can present this correction as follows
$$
\delta \bm s_{\bm k}^{(j)} = - F_{k}^{(j)} \left( \bar{\bm s}_{\bm k}^{(j)} \times \bm \Omega_{\bm
k}^{(j)}\right) \:,
$$
where $F_{k}^{(j)}$ is a function of $k = |\bm k|$. It can be
found from the solution of linearized Eq.~\eqref{sk:gen}. For the
Rashba-like spin splitting we eventually arrive at
\begin{equation}
\frac{1}{\tau_{s,zz}^{(j)}} = \sum_{\bm k} {\Omega^{(j)}_{\bm k}}^2 F_{k}^{(j)}
= \beta_j^2  \sum_{\bm k} k^2 F_{k}^{(j)}, \label{tauzz}
\end{equation}
and $\tau_{s,xx}^{(j)} = \tau_{s,yy}^{(j)} = 2 \tau_{s,zz}^{(j)}$,
where $\tau_{s, \alpha \alpha}^{(j)}$ is the spin relaxation time
in the $j$-th subband for the spin oriented along the $\alpha$
axis.

In the limits of degenerate and non-degenerate statistics it is
instructive to introduce an effective scattering time $\tau_j^*$
in the $j$th subband defined by
\begin{equation}
\label{tauzz1}
\frac{1}{\tau_{s, zz}^{(j)}} = \Omega_j^2 \tau_j^*\:,
\end{equation}
where the characteristic spin precession frequency $\Omega_j =
\beta_j k_F^{(j)}$ for a degenerate electron gas and $\Omega_j =
\beta_j k_T$ for a non-degenerate gas, $k_F^{(j)}$ is the Fermi
wavevector at zero temperature in a given subband, and $k_T$ is
the thermal wavevector $\sqrt{2m^* k_B T}/\hbar$. In fact, the time
$\tau_j^*$ is a microscopic electron-electron scattering time
governing the Dyakonov-Perel' spin relaxation in each subband.
Comparing Eqs.~\eqref{tauzz} and \eqref{tauzz1} we obtain
\begin{eqnarray}
 \tau^*_j = \sum_{\bm k} \frac{k^2}{{k_F^{(j)}}^2} F_k^{(j)} \hspace{8 mm} \mbox{(degenerate
electrons)}\: , \label{star:de} \\
 \tau^*_j = \sum_{\bm k} \frac{k^2}{k_T^2} F_k^{(j)} \hspace{5 mm} \mbox{(non-degenerate
electrons)}\:.
\end{eqnarray}

\section{Results and discussion}

Below we present analytical and numerical results for the
microscopic scattering times $\tau_j^*$ which govern
Dyakonov-Perel' spin relaxation in multivalley QWs. In order to
emphasize the role of electron-electron interaction the effects of
single-particle momentum scattering are ignored, they can be taken
into account by inclusion into the right-hand side of kinetic
equation~\eqref{sk:gen} the collision term $- \delta \bm
s_{\bm k}/\tau_p$, where $\tau_p$ is the momentum scattering time.

% In the analytical treatment we neglect the terms $V_{\bm k - {\bm
% p}} V_{\bm k - {\bm p}'}$ in the collision integrals since they
% make minor influence on spin relaxation
% times~\cite{glazov02,glazov04a}. Then, 

For the
\emph{non-degenerate} electron gas, one can neglect the screening of the electron-electron interaction and the collision integrals describing intra-subband and
subband-subband $\bm Q_{\bm k}\{\bm s,f\}$ and
$\tilde{\bm Q}_{\bm k}\{\bm s,f\}$ differ, apart from the terms $V_{\bm k - {\bm
p}} V_{\bm k - {\bm p}'}$, by a common factor, $f_{\bm k'}^{(j)}/f_{\bm k'}^{(-j)}$, resulting from different populations of the valley-orbit split subbands.
The inverse microscopic scattering time ${\tau^*}^{-1}$ has two
additive contributions caused by the collision of electrons within
the same subband and electrons in different subbands each of those being proportional to the number of electrons in a given subband. Neglecting the terms $V_{\bm k - {\bm
p}} V_{\bm k - {\bm p}'}$ and making use
of the results for a single valley~\cite{glazov02,glazov04a} we
have
\begin{equation}
 \label{tau:nondeg}
\tau^*_- = \tau^*_+ =  \tau^{(B)}_{ee} \:,
\end{equation}
where $\tau_{ee}^{(B)}$ is the electron-electron scattering time which
governs spin relaxation in the single valley structure occupied by
electrons with total concentration $N = N_+ + N_-$,
\begin{equation}
 \label{tau:0:nondeg}
\tau_{ee}^{(B)} = \frac{\hbar\ae^2 k_B T}{e^4 N} I\:,
\end{equation}
and $I$ is a numerical factor which, for strictly two-dimensional
electrons, equals to $\approx 0.027$ ~\cite{glazov02,glazov04a}.
The scattering times in the subbands are, therefore, the same, since it does not matter whether an electron scatters by an electron in the same or in the other subband. The spin relaxation times in the valley-orbit split subbands are different only due to the difference of the spin splittings in the subbands. Note, that the allowance for the interference contributions $V_{\bm k - {\bm p}} V_{\bm k - {\bm p}'}$ results in a slight ($\approx 4\%$) increase of the constant $I$ for the intrasubband interaction~\cite{glazov02,glazov04a}, hence, these terms can be safely neglected.

\begin{figure}
% \centerline{\includegraphics[width=0.35\textwidth]{nd1.eps}}
\centerline{\includegraphics[width=0.35\textwidth]{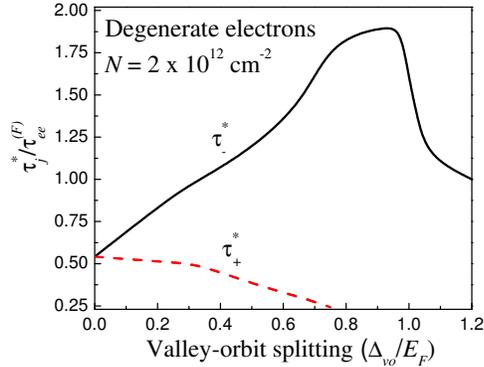}}
\caption{Electron-electron scattering times $\tau^*_{\pm}$ in a two-valley QW
as a function of the valley-orbit splitting. The times are presented
in units of the similar scattering time in a single valley with the same
carrier density and temperature. Black solid line corresponds to the lower, more populated, valley
while red dashed line describes the upper, less populated, valley. The calculation is carried out for the degenerate electron gas, temperature $T=8.2$~K,
$N=2 \times 10^{12}$~cm$^{-2}$, $E_F$ is the Fermi energy in the single
valley with the same concentration $N$. Other parameters used
in the calculation correspond to Si/SiGe QWs: $\ae=12$, and $m^* = 0.191 m_0$, where $m_0$ is
the free electron mass. }
\label{fig:rates}
\end{figure}

Now we turn to low temperatures where the electrons are
\emph{degenerate}. Figure~\ref{fig:rates} depicts the dependence
of the scattering times $\tau^*_{\pm}$ on $\Delta_{\rm vo}$
related to the Fermi energy $E_F$ of electrons of the same
concentration populating a single valley. In this case the
electron-electron collisions are suppressed due to the Pauli
principle and, moreover, the screening parameter $q_s$ is not
negligible. This gives rise to two additional competing factors
which have effect on the difference between the scattering rates
in single- and two-valley systems. First, the electrons are
redistributed between valleys which results in a decrease of
electron concentration in each valley and, consequently, in an
enhancement of the scattering rate due to reducing the Pauli
blocking. Second, the screening efficiency increases and,
therefore, the scattering rates are decreased. Due to the
competition between these two factors the electron-electron
scattering time can be both longer and shorter in a two-valley
system as compared with a single valley.

A simple analytical expression for the electron-electron scattering
time can be derived in the absence of intervalley mixing in which case $N_+ = N_- = N/2$, $\mu_+= \mu_- = E_F/2$ and $\tau^*_+ = \tau^*_- \equiv \tau^*$. Let us introduce the electron-electron scattering rate governing the spin relaxation in a single-valley system with the degenerate electrons of the total density $N$~\cite{glazov04a}
\begin{equation}
 \label{tau:F}
 \frac{1}{\tau_{ee}^{(F)}} \approx 3.4 \frac{(k_B T)^2}{ \hbar E_F } = 3.4 \frac{m (k_B T)^2}{\pi \hbar^3 N}\:.
\end{equation}
In a two-valley system at, $k_B T \ll \mu$, one has
\begin{equation} \label{tau:deg}
\tau^* = \frac{J}{4} \tau_{ee}^{(F)}\:.
\end{equation}
The factor $1/4$ in Eq.~\eqref{tau:deg} results from allowance for the valley-valley interaction and takes also into account that the Fermi energy in each valley is twice smaller as compared with the single-valley system with the same total density. The factor $J$ describes the
modification due to allowance for the screening. In the limit of
$k_F = \sqrt{m^* E_F/\hbar^2} \ll q_s$, i.e., where the
screening is so strong that the electron-electron interaction is
effectively short-range, $J = 4$ since the inverse screening
length is twice smaller as compared with single valley system and,
hence, the scattering probability decreases by a factor of 4. In
real QWs $k_F$ and $q_s$ can be comparable~\cite{PhysRevB.53.7403}
and $J$ ranges from $1$ to $4$ depending on the electron
concentration. For the parameters used in the calculation of
Fig.~\ref{fig:rates} the factor $J \approx 2.2$ and, at
$\Delta_{\rm vo} \ll E_F$, the ratio $\tau^*/\tau_{ee}^{(F)}$ is
close to 0.55. In QW structures with a large number of unmixed
valleys, $n_v \gg 1$, like (111)-grown Si MOSFET structures and degenerate electrons, the scattering
time $\tau^*$ increases $\propto n_v$ due to the competing effects
of enhancing screening and decreasing Pauli blocking. With the
increasing valley-orbit splitting, the electron-electron
scattering time in the lower valley, $\tau_-$, becomes longer and the
scattering time $\tau_+$ shortens. This is a result of
electron redistribution downward to the lower subband and an
enhancement of Pauli blocking there. In the upper subband the
electron density decreases and the Pauli blocking becomes weaker.
If all the electrons fill the lower subband the scattering time,
$\tau_-$, rapidly drops because the screening parameter $q_s$
reduces by a factor of 2 and approaches the single-valley value,
Fig.~\ref{fig:rates}. One can also see from this figure that
the electron-electron scattering time $\tau^*_-$ can be both
shorter and longer than that for the single-valley
system.

For non-zero valley-orbit splitting, the spin relaxation times of the electrons in the two subbands $j=\pm$ can be different due to the following reasons: (i) difference of the electron-electron scattering times $\tau_+^*\ne \tau_-^*$, (ii) difference of the Fermi wavevectors $k_F^{(j)}$ and (iii) difference of the spin-splitting constants $\beta_+ \ne \beta_-$. Weak intervalley scattering characterized by the time $\tau_v \gg \tau_{\pm}^*, \tau_p$ may lead to the efficient intermixing of spins in different valleys. The observed spin relaxation time for the spin along one of the main axes $\alpha$ is, hence,
$$T_{s,\alpha\alpha} = \frac{2\tau_{s,\alpha\alpha}^{(+)}\tau_{s,\alpha\alpha}^{(-)}}{\tau_{s,\alpha\alpha}^{(+)}+ \tau_{s,\alpha\alpha}^{(-)}},$$
provided $\tau_v \ll \tau_{s, \alpha\alpha}^{(\pm)}$.

\begin{figure}
%\centerline{\includegraphics[width=0.35\textwidth]{tdep1.eps}}
\centerline{\includegraphics[width=0.35\textwidth]{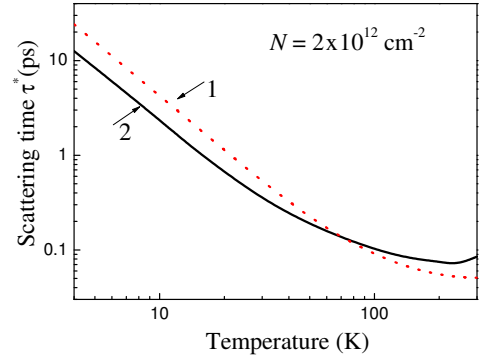}}
\caption{Electron-electron scattering times as a function of temperature
calculated from Eq.~\eqref{star:de} for single valley (curve 1) and two-valley
(curve 2) quantum wells. The valley-orbit splitting is set to zero. The electron
concentration $N=2\times 10^{12}$ cm$^2$. Other parameters are the same as in
caption to Fig.~\ref{fig:rates}.}
\label{fig:tdep}
\end{figure}

Finally, in Fig.~\ref{fig:tdep} the calculated temperature dependence of the
electron-electron scattering time is depicted. Dotted curve
represents a single-valley system, solid curve shows the
calculation for the two-valley QW with zero valley-orbit splitting and the same concentration of
carriers. We remind that according
to Eq.~(\ref{tauzz1}) the spin relaxation rate is obtained as a
product of $\tau^*$ defined by Eq.~\eqref{star:de} and the squared
spin precession frequency taken at the Fermi level at zero
temperature. The qualitative behavior of these two curves is
similar: with the temperature increase the scattering time
shortens as $\tau^* \propto T^{-2}$~[see Eq.~\eqref{tau:deg}] due
to the weakening of Pauli blocking and reaches a minimum (seen in
the figure only for the two-valley structure) caused by the
transition to the non-degenerate case. This transition takes place
at a smaller temperature for the two-valley system because the
carrier concentration in each valley is twice smaller. For the
accepted parameters the scattering time in the two-valley system,
in comparison with the single-valley system, is shorter at lower
temperatures and longer at higher temperatures.

One can see from Fig.~\ref{fig:tdep} that the scattering time
$\tau^*$ has a picosecond scale in a wide range of temperatures.
In the state-of-the-art Si/SiGe QWs where the spin relaxation was
studied the momentum scattering time $\tau_p$ was about $10$~ps
for even smaller carrier concentrations than those taken in our
calculation. Therefore, electron-electron collisions play a
substantial role in controlling the spin relaxation in those
Si/SiGe structures.

\section{Conclusions}

We have developed a theory of electron-electron scattering effect
on the Dyakonov-Perel' spin relaxation in multi-valley
semiconductor QWs. We have shown that, although the intervalley scattering
of electrons is suppressed, the interaction of electrons occupying 
different valley-orbit split subbands influences the spin relaxation. The
electron-electron scattering rates in single and multi-valley
systems are different due to (i) redistribution of electrons between the subbands, and (ii) an
enhancement of screening in the two-valley systems.

The values of electron-electron scattering times in high-mobility
Si/SiGe QWs with two occupied valleys may be comparable and even
shorter than the momentum scattering time in a wide range of
temperatures. Therefore, in these structures the electron spin
relaxation can be controlled by electron-electron scattering.

\acknowledgments Authors thank Ming-Wei Wu, W. Jantsch and Z. Wilamowski for
valuable discussions and M.O. Nestoklon for critical reading of the manuscript. We gratefully acknowledge the financial
support from RFBR, Programs of RAS and ``Dynasty'' Foundation
--- ICFPM.

% \bibliographystyle{eplbib}
% \bibliography{/home/misha/Work/Coherent/Bibliography/all}

\end{document}